\documentclass[epj]{webofc}\usepackage[varg]{txfonts}
\woctitle{QCD@Work 2016}
\begin{document} % HEPHY-PUB 972/16 (2016)
\title{On properties of the exotic hadrons from QCD sum rules}
\author{Wolfgang Lucha
\inst{1}\fnsep\thanks{\email{Wolfgang.Lucha@oeaw.ac.at}}\and
Dmitri Melikhov
\inst{1,2,3}\fnsep\thanks{\email{dmitri_melikhov@gmx.de}}}
\institute{Institute for High Energy Physics, Austrian Academy of
Sciences, Nikolsdorfergasse 18, A-1050 Vienna, Austria \and
D.~V.~Skobeltsyn Institute of Nuclear Physics, M.~V.~Lomonosov
Moscow State University, 119991, Moscow, Russia \and Faculty of
Physics, University of Vienna, Boltzmanngasse 5, A-1090 Vienna,
Austria}\abstract{We discuss the specific features of extracting
properties of the exotic polyquark hadrons (tetraquarks,
pentaquarks) compared to the usual hadrons by the QCD sum-rule
approach. In the case of the ordinary hadrons, already the
one-loop leading-order $O(\alpha_s^0)$ correlation functions
provide the bulk of the hadron observables, e.g., of the form
factor; inclusion of radiative corrections $O(\alpha_s)$ modifies
already nonzero one-loop contributions. In the case of an exotic
hadron, the situation is qualitatively different: discussing
strong decays of an exotic tetraquark meson, which provide the
main contribution to its width, we show that the {\it
disconnected} leading-order diagrams are not related to the
tetraquark properties. For a proper description of the tetraquark
decay width, it is mandatory to calculate specific radiative
corrections which generate the {\it connected} diagrams.}
\maketitle

\section{Properties of individual resonances from QCD correlation
functions}One of the most famous applications of the method of QCD
sum rules (see \cite{svz,colangelo,ioffe}) is the calculation of
the hadronic ground-state properties from the QCD correlation
functions involving power corrections.

\subsection{Two-point functions}To recall the basics of the method,
we briefly review QCD sum rules arising from two-point functions.
For obtaining properties of a hadron $M$, one considers an
interpolating current for this hadron, i.e., a current $j(x)$
which produces this hadron from the vacuum:\begin{eqnarray}
\label{f}\langle\Omega|j(0)|M\rangle=f_M\ne 0.\end{eqnarray} (For
instance, $j(x)=\,\bar q_1(x)O q_2(x)$ for ``normal'' mesons,
four-quark currents for exotic mesons). One then considers the
correlation functions, i.e., the vacuum expectation values of the
$T$-products of the interpolating currents. The simplest object is
the two-point function
\begin{eqnarray}\Pi(p^2)=i\int d^4x\,e^{ipx}\left\langle\Omega
\left|T\left(j(x)j^\dag(0)\right)\right|\Omega\right\rangle.
\end{eqnarray}Wilson's operator product expansion (OPE) provides
the following expansions, for the $T$-product \cite{nsvz1984},
\begin{eqnarray}T\left(j(x)j^\dagger(0)\right)=C_0(x^2,\mu)\hat1+
\sum\limits_n C_n(x^2,\mu) :\!\hat O_n(x=0,\mu)\!:,\end{eqnarray}
and, for the two-point function,\begin{eqnarray}\Pi(p^2)=\Pi_{\rm
pert}(p^2,\mu)+\sum_n \frac{C_n}{(p^2)^n}\langle\Omega|:\!\hat
O_n(x=0,\mu)\!:|\Omega\rangle.\end{eqnarray}The basic concept of
the method is the difference between the physical QCD vacuum,
$|\Omega\rangle$, which has a complicated structure, and the
perturbative QCD vacuum, $|0\rangle$: properties of the physical
vacuum~$|\Omega\rangle$ are characterized by the condensates ---
nonzero expectation values of gauge-invariant operators over the
physical vacuum: $\langle\Omega|:\!\hat O(0,\mu)\!:|\Omega\rangle
\ne0$ (see \cite{condensate_qq,condensate_GG} for the recent
determinations). Hereafter,~for notational simplicity, we write
$\langle\Omega|\cdots|\Omega\rangle\equiv\langle\cdots\rangle$.
The two-point function is an analytic function~of~$p^2$:
\begin{eqnarray}\Pi(p^2)=\int \frac{ds}{s-p^2}\rho(s).\end{eqnarray}
One calculates the spectral densities $\rho(s)$ using both OPE and
the language of confined hadron states:\begin{eqnarray}\rho_{\rm
theor}(s)=\left[\rho_{\rm pert}(s,\mu)+\sum_n C_n
\delta^{(n)}(s)\langle\Omega|\hat O_n(\mu)|\Omega\rangle\right],
\qquad\rho_{\rm hadr}(s)=f^2\delta(s-M^2)+\rho_{\rm cont}(s).
\end{eqnarray}In order to relate to each other the truncated
$\Pi_{\rm OPE}(p^2)$ and the hadronic representation $\Pi_{\rm
hadron}(p^2)$, one has to perform a smearing, e.g., by application
of a Borel transformation $p^2\to \tau$ $\left[\frac{1}{s-p^2}\to
\exp(-s\tau)\right]$:\begin{align}\Pi(\tau)=\int ds\,e^{-s
\tau}\rho(s)&=f^2 e^{-M^2\tau}+\int\limits_{s_{\rm phys}}^\infty
ds\,e^{-s\tau}\rho_{\rm hadr}(s)\nonumber\\&=
\int\limits_{(m_1+m_2)^2}^\infty ds\,e^{-s\tau}\rho_{\rm
pert}(s,\mu)+\Pi_{\rm power}(\tau,\mu).\end{align}Here, $s_{\rm
phys}$ is the physical threshold, and $f$ is the decay constant
defined by (\ref{f}). In order to get rid of the excited-state
contributions, one adopts the {\it duality ansatz}
\cite{shifman1,lm,qcdvsqm}: all contributions of excited states
are counterbalanced by the perturbative contribution above some
{\em effective continuum threshold}~$s_{\rm eff}(\tau),$ which
differs from the physical continuum threshold. Applying the
duality assumption yields\begin{eqnarray}\label{2ptsr}f^2
e^{-M^2\tau}=\int\limits_{(m_1+m_2)^2}^{s_{\rm
eff}(\tau)}ds\,e^{-s\tau}\rho_{\rm pert}(s,\mu)+\Pi_{\rm
power}(\tau,\mu).\end{eqnarray}As soon as we have an algorithm
\cite{lms_2ptsr,lms_new} of fixing the effective threshold $s_{\rm
eff}(\tau)$ at our disposal,~Eq.~(\ref{2ptsr}) provides the decay
constant $f$. To this end, we need either to know the hadron mass
$M$ or to set~$\tau=0$.

\section{Strong decays from three-point functions in QCD}Having
recalled these basic ideas, let us now consider three-point
functions, which are the appropriate QCD quantities for various
hadron form factors \cite{ioffe3pt,radyushkin}:\begin{eqnarray}
\label{3pt}\Gamma(p,p',q)=\int dx\,dx'\,\langle T(J(x)j(0)j'(x')
\rangle\exp(ipx-ip'x').\end{eqnarray}

\begin{figure}[t]\centering\sidecaption$\qquad\qquad\qquad\qquad$
\includegraphics[width=3cm,clip]{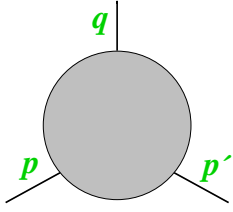}\caption{Three-point
vertex function (\ref{3pt}).}\label{Fig:3pt}\end{figure}This
correlator contains the triple-pole in the Minkowski region,
namely,\begin{eqnarray}\Gamma(p,p',q)=\frac{f
f'}{(p^2-M^2)({p'}^2-{M'}^2)}F(q^2)+\cdots,\end{eqnarray}where the
form factor $F(q^2)$ contains a pole at $q^2=M_q^2$:
\begin{eqnarray}F(q^2)=\frac{f_qg_{MM'M_q}}{(q^2-M_q^2)}+\cdots.
\end{eqnarray}Here, $g_{MM'M_q}$ describes the $M\to M'M_q$ strong
transition; $f$, $f'$, and $f_{M_q}$ are the decay constants
of~the mesons: $\langle0|J(0)|M\rangle=f$,
$\langle0|j'(0)|M'\rangle=f'$, and $\langle0|j(0)|M_q\rangle=f_q$.

The three-point function satisfies the double spectral representation
\begin{eqnarray}\Gamma(p,p',q)=\int\frac{ds}{s-p^2}\frac{ds'}{s'-{p'}^2}
\Delta(s,s',q^2)\end{eqnarray}A double Borel transformation
$p^2\to\tau$, $p'^2\to\tau'$ and duality cuts in the $p^2$ and
$p'^2$ channels \cite{ioffe3pt,radyushkin,bakulev} lead to
\begin{eqnarray}\label{3ptsr}\exp(-M^2\tau)\exp(-M'^2\tau')F(q^2)
=\int\limits^{s_{\rm eff}}ds\exp(-s\tau)\int\limits^{s'_{\rm eff}}
ds'\exp(-s'\tau')\Delta_{\rm OPE}(s,s',q^2)\end{eqnarray}Let us
discuss this relation for two cases: for normal bilinear and for
exotic four-quark currents.

\subsection{Normal hadrons}An elastic or a transition form factor
of a normal meson corresponds to the choice of the interpolating
currents in Eq.~(\ref{3ptsr}) in the bilinear form, that is,
$J,j,j'=\bar q_i\hat Oq_j$, with appropriate quark fields
$q_{i,j}$ and a generic Dirac matrix $\hat O$. The perturbative
part of the correlation function has the following~structure:
\begin{eqnarray}\Gamma_{\rm OPE}(p^2,p'^2,q^2)=
\Gamma_0(p^2,p'^2,q^2)+\alpha_s\Gamma_1(p^2,p'^2,q^2)+\cdots.
\end{eqnarray}

\begin{figure}[h]\centering\sidecaption
\includegraphics[width=8.8cm]{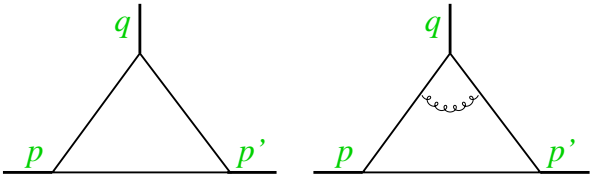}\caption{The OPE for
the three-point function of the bilinear $\bar q q$ interpolating
currents for normal mesons. The leading-order diagram and one of
the radiative corrections are shown.}\label{Fig:normal}
\end{figure}Already the one-loop leading-order diagram has a
nonzero double-spectral density $\Delta_0(s,s',q^2)$ and therefore
provides a nonzero contribution to the form factor (and,
respectively, to $g_{MM'M_q}$). Radiative corrections
\cite{bagan,fulvia,braguta} give essential contributions,
increasingly important at larger $q^2$
\cite{anisovich,blm2008plb}. However, a reasonable estimate may be
obtained already from the leading-order three-point function
\cite{lm2012prd,lm2012jpg,blm2012prd,ms2012prd}.

\subsection{Three-point function containing one exotic current}
\begin{figure}[h]\centering\sidecaption
\includegraphics[width=8.8cm,clip]{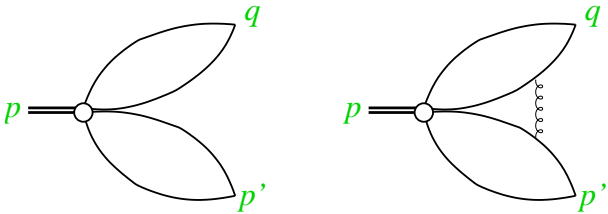}
\caption{Perturbative expansion of the three-point function
containing one exotic tetraquark current and two normal
quark--antiquark currents.} \label{Fig:exotic}\end{figure}

The
interpolating current for the exotic state (for the sake of
clarity, we focus to the case of tetraquarks, i.e., exotic states
with four valence quarks) may be chosen in many different ways
(see the discussion in the next section). Let us consider an
interpolating current $\theta(x)$ of the following rather general
form:\begin{eqnarray}\theta(x)=\bar q_1(x)\hat Oq_2(x)\bar q_3(x)
\hat Oq_4(x),\end{eqnarray}where $\hat O$ is an appropriate
combination of Dirac and colour matrices and possibly also of
(covariant) derivatives. The OPE for the correlation function
$\Gamma^{\theta jj}=\langle T(\theta(x)j(y)j(z)\rangle$ containing
one exotic current and two normal currents has the diagrammatic
structure shown in Fig.~\ref{Fig:exotic} and reads\begin{eqnarray}
\Gamma^{\theta jj}_{\rm OPE}(p^2,p'^2,q^2)=\Pi(p'^2)\Pi(q^2)
+\alpha_s\Gamma_{\rm connected}(p^2,p'^2,q^2).\end{eqnarray}

As known from the general features of the Bethe-Salpeter equation,
the disconnected diagrams~are not related to the bound states.
This is also clear from the following argument: Performing the
Borel transformation $p^2\to\tau$ (one of the steps of a sum-rule
analysis), the Borel image of the disconnected leading-order
contribution vanishes. The {\it connected} diagrams, relevant for
the exotic states, emerge~at order $O(\alpha_s)$ and higher. Thus,
any attempt to extract, e.g., the tetraquark decay amplitude from
the leading-order contribution is inconsistent. The common feature
of {\it all} previous investigations of these decays by QCD sum
rules (e.g., \cite{nielsen2010,nielsen2014,wang2014,nielsen2016})
was the attempt to study tetraquark (and pentaquark) decays
relying on the factorizable leading-order contribution, which
intrinsically has no relation to tetraquark properties [which fact
is evident from both the factorization property $\Gamma(p,p',q)=
\Pi(p'^2)\Pi(q^2)$ and the large-$N_c$ behaviour of the QCD
diagrams \cite{Weinberg,knecht,cohen}]. Therefore, the existing
analyses should be strongly revised by calculating and taking into
account the nonfactorizable two-loop $O(\alpha_s)$ corrections. In
other words, {\it the ``fall-apart'' decay of exotic hadrons
differs from the decay mechanism of ordinary hadrons and requires
an appropriate treatment within QCD sum rules. The leading-order
contribution is not related to the exotic-state decay, as is clear
from its factorization property $\Gamma(p,p',q)=\Pi(p'^2)\Pi(q^2)$
and from the large-$N_c$ behaviour of the QCD diagrams. The
calculation of the radiative corrections is mandatory for a
reliable analysis of the properties of the exotic states.}

\section{Structure of the exotic states}Obviously, the exotic
tetraquark states may have a rather complicated ``internal''
structure. The most popular scenarios for such a structure are a
confined tetraquark state (i.e., a bound state in a confining
potential between two colour-triplet diquarks) and a molecular
``nuclear-physics-like'' bound state in the system of two
colourless mesons.

However, an important question about the structure of the exotic
state --- which to a large extent controls also its production
mechanism --- is not easy to answer \cite{jaffe}: (i) by a
combined colour--spinor Fierz rearrangement of the tetraquark
interpolating current $D(x)$ one can express it in either
diquark--antidiquark or meson--meson form; (ii) the same quantum
numbers of the exotic interpolating current may be obtained by
different combinations of its diquark--antidiquark or meson--meson
bilinear parts.

\subsection{Set of decay constants of exotic states}The simplest
characteristic of a meson is its decay constant, i.e., the
transition amplitude between~the vacuum and the meson, induced by
its interpolating current. For a heavy quarkonium state, the decay
constant is analogous to its wave function at spatial origin,
$\psi(r=0)$. Whereas an ordinary meson is described by one or few
decay constants $\langle0|\bar q_1O_Aq_2|M\rangle=f_A$, exotic
states have many decay constants, related to possible different
structures of the interpolating four-quark currents; e.g., the
interpolating currents may be chosen in ``meson--meson'' form
$\langle0|M_{12}(x)M_{34}(x)|\theta\rangle=f_{MM}$ with
$M_{12}=\bar q_1Oq_2$,~or~in ``diquark--diquark'' form
$\langle0|\bar D^{a}_{13}(x)D^{a}_{24}(x)|\theta\rangle=f_{DD}$
with $D^{a}_{24}=\epsilon^{abd}(q^b_{c4})^Tq^d_2.$ If, in some
limit, a certain class of decay constants vanishes, one is
eligible to say that the exotic state is, e.g., a pure
molecular~or a pure diquark--antidiquark state. In general, the
mixing of different ``components'' seems unavoidable.

An appropriate tool to access the decay constants theoretically is
the QCD sum rule for two-point functions,
$\Pi^{(\theta\theta)}=\langle\theta(x)\theta(0)\rangle$
\cite{narison}. Let us emphasize, however, that {\it not all
contributions to two-point functions of exotic interpolating
currents are related to the properties of exotic states. The
large-$N_c$ behaviour of the different diagrams should be the
guiding principle for the selection of appropriate contributions,
similar to selecting the connected contributions to three-point
functions}. After selecting the appropriate contributions to the
two-point functions $\Pi_{\theta \theta}$, the set of sum rules
should be studied and only then the answer about the structure of
the observed narrow exotic candidates may be obtained.

\subsection{Form factors of the exotic states}The appropriate
quantities which describe the structure of bound states --- both
normal and exotic --- are the form factors. For exotic states,
these are the form factors related to the connected parts of the
three-point functions involving two exotic interpolating currents,
$\langle T(\theta\theta j)\rangle$; so far, no OPE calculations
for the corresponding quantities are available in the literature.

\section{Conclusions and Outlook}\begin{itemize}\item The
three-point (vertex) correlation functions for the normal bilinear
quark currents show a crucial difference from the three-point
correlation functions of exotic polyquark interpolating
currents:~the latter are described at leading order by
disconnected diagrams, not related to strong decays of exotic
hadrons. In the exotic-current case, the relevant diagrams start
at order $O(\alpha_s)$. This requires taking into account
$O(\alpha_s)$ corrections.\item Similar to the case of three-point
functions, not all contributions to two-point correlation
functions of exotic interpolating currents are related to the
properties of the exotic bound state: the $N_c$ scaling may be the
guiding principle for selecting the appropriate contributions.
\item Our experience in the analysis of the normal hadrons proves
that a truncated OPE for the correlation function does not enable
one to study at the same time both the {\it existence} of the
isolated ground~state and its {\it properties}. However, if the
mass of such a narrow bound state is known, the QCD-sum rule
approach allows one to obtain reliable predictions for decay
constants \cite{lms_fBetc} and form factors \cite{hagop}. The
observed exotic-state candidates are narrow; therefore, any
procedure for the extraction of their parameters from the OPE has
the same features and the same challenges as for the normal
hadrons.\end{itemize}

\acknowledgement D.~M. was supported by the Austrian Science Fund
(FWF) under project P29028-N27.

\bibliographystyle{aipproc}
\end{document}